\documentclass[aps,prl,twocolumn,showpacs]{revtex4-1}  
\usepackage{amsmath}
\usepackage{amssymb}
\usepackage{graphicx}
\usepackage{epstopdf}

\begin{document}

\title{Quantum-impurity relaxometry of magnetization dynamics}

\author{B. Flebus}
\author{Y. Tserkovnyak}
\affiliation{Department of Physics and Astronomy, University of California, Los Angeles, California 90095, USA}

\begin{abstract}
Prototypes of quantum impurities (QI), such as NV and SiV centers in diamond, have been recently growing in popularity due to their minimally invasive and high-resolution magnetic field sensing. Here, we focus on quantum-impurity relaxometry as a method to probe collective excitations in magnetic insulators. We develop a general framework that relates the experimentally-measurable quantum-impurity relaxation rates to the properties of a magnetic system via the noise emitted by the latter. We suggest that,  when the quantum-impurity frequency lies within the spin-wave gap, quantum-impurity relaxometry can be effectively deployed to detect signatures of the coherent spin dynamics, such as magnon condensation,  both in ferromagnetic and antiferromagnetic systems, as well as open prospects to nonintrusively probe spin-wave transport regimes in magnetic insulators.
\end{abstract}


%75.76.+j		Spin transport effects
%72.25.Mk	Spin transport through interfaces
%72.20.Pa	Thermoelectric and thermomagnetic effects
%85.75.-d		Magnetoelectronics; spintronics: devices exploiting spin polarized transport or integrated magnetic fields

\maketitle

\textit{Introduction}. Quantum impurities (QI), such as NV and SiV centers in diamond, can display an exceptional sensitivity to magnetic fields and their spin state can be initialized and read out optically \cite{Taylor2008,Pham2011, Hong2013,Grinolds2013}. These properties make them ideal candidates for probing the magnetic or electronic noise emitted by a nearby system via their spin relaxation rates. The noise can then be related, via fluctuation-dissipation theorem~\cite{kubo1966}, to the response functions, and, hence, the physical properties of the system under investigation.
While quantum-impurity relaxometry has been already proposed as a platform to study transport properties and spatial inhomogeneities in electronic systems~\cite{Agarwal2017}, an analogous theoretical framework for magnetic insulators is still partially lacking. Nonetheless, the quantum-impurity ability to probe noise locally and nonintrusively appears particularly promising for magnetic insulating systems~\cite{Du2017}, as the detection of their collective excitations, i.e., spin waves, is otherwise largely limited to conventional spin transport experiments \cite{CornelissenNature2015,CornelissenPRB2016} or microwave probes~\cite{ChumakAPL2012}.

Spin-wave relaxometry has been heretofore focusing only on the noise emitted by a magnetic system at  frequencies higher than its spin-wave gap~\cite{DerSar2015,Du2017}. This noise reflects both the spectrum and the distribution of the magnon gas at the  QI resonance frequency~\cite{Du2017}.   Following this approach, Du \textit{et al.} have provided the first direct measurement of the magnon chemical potential, as well as its dependence on  external perturbations, in a ferromagnetic system \cite{Du2017}. However, a variety of magnetic systems possess gaps that are exceedingly large for microwave excitations. In this Letter, therefore, we address the detection, via quantum-impurity relaxometry, of the magnetic noise emerging at subgap frequencies. 

%We find that this noise, until now neglected, not only can signal dynamical phase transition such as Bose-Einstein condensation of magnons, but it can be deployed to directly probe spin-wave transport properties.

\begin{figure}[b!]
\includegraphics[width=1\linewidth]{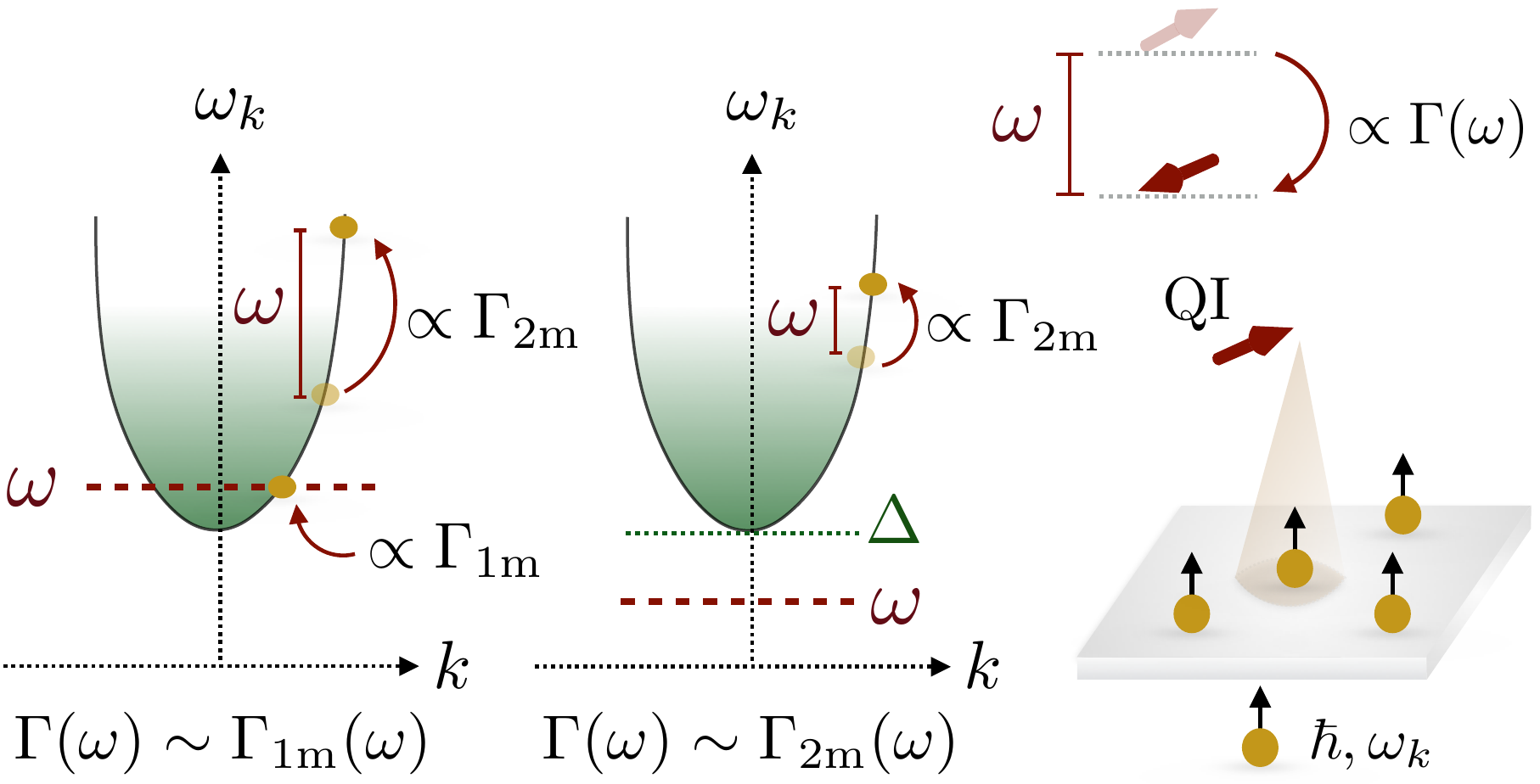}
\caption{Quantum-impurity relaxation via one- and two-magnon processes. The interaction between the QI spin and a nearby magnetic system, here depicted as  gas of magnons with spin  $\hbar$ and frequency $\omega_{k}$ (with $\omega_{k=0}=\Delta$), leads  to a QI transition with emission of energy $\hbar \omega$. When $\omega>\Delta$, the latter can result in the creation of a magnon at frequency $\omega_{k}=\omega$ or in a magnon scattering with energy gain $\hbar \omega$. These events contribute, respectively, to the single-magnon, $\Gamma_{\text{1m}}$, and two-magnon, $\Gamma_{\text{2m}}$, relaxation rate. When $\omega>\Delta$, the relaxation rate is typically dominated by one-magnon processes. Conversely, for $\omega<\Delta$, one-magnon events are suppressed and $\Gamma \sim \Gamma_{\text{2m}}$. The overall relaxation rate accounts as well for QI transitions with absorption of energy $\hbar \omega$.}
\label{fig1NV}
\end{figure}

The interaction between spin waves and a QI spin induces QI transitions between its spin states. When the QI spin relaxes, it releases energy proportional to its resonance frequency. How this energy is converted into excitations of the magnetic system depends on the gap  of the spin-wave spectrum. If the QI resonance frequency is larger than the magnetic gap,  QI relaxation can trigger both one- and two- magnon processes, corresponding, respectively, to the creation of a magnon at the QI resonance frequency or to a magnon scattering with energy gain equal to it, as depicted in Fig.~\ref{fig1NV}.  One can show, already by using a simple model of a local interaction between the QI spin and the ferromagnetic spin density of an ideal  magnon gas, that the relaxation rate, $\Gamma_{2m}$, due to two-magnon processes is suppressed at low temperatures with respect to the one-magnon, $\Gamma_{1m}$, one, i.e., $\Gamma_{1m} \sim (T/T_{C})$ and  $\Gamma_{2m} \sim (T/T_{C})^2$, where $T_{C}$ is the Curie temperature.
However, when the QI resonance frequency lies within the gap, one-magnon scattering is prohibited 
and two-magnon processes overtake the quantum-impurity transitions. Focusing on this regime,
 we develop a  theory of quantum-impurity relaxometry driven by two-magnon noise.  To illustrate its capability of probing spin transport properties and detecting dynamic phase transitions, we discuss  two main examples. First, we show that two-magnon driven  QI relaxometry can directly probe diffusive spin-wave bulk transport properties,  which cannot be easily extracted from the one-magnon noise and whose manifestations in the state-of-the-art spin-transport experiments  are intertwined with spin conversion at the insulator$|$metal interfaces~\cite{CornelissenNature2015,CornelissenPRB2016}.  Finally, we investigate the dependence of the two-magnon noise on the magnon chemical potential in both ferromagnetic and antiferromagnetic systems. We find that the two-magnon noise alone can signal the precipitation of a Bose-Einstein condensation.

%Finally, we note that characterizing the coupling between quantum impurities and \textit{U}(1)-symmetric magnets opens up new perspectives to couple distant qubits by magnetic materials.

\begin{figure}[t!]
\includegraphics[width=0.70\linewidth]{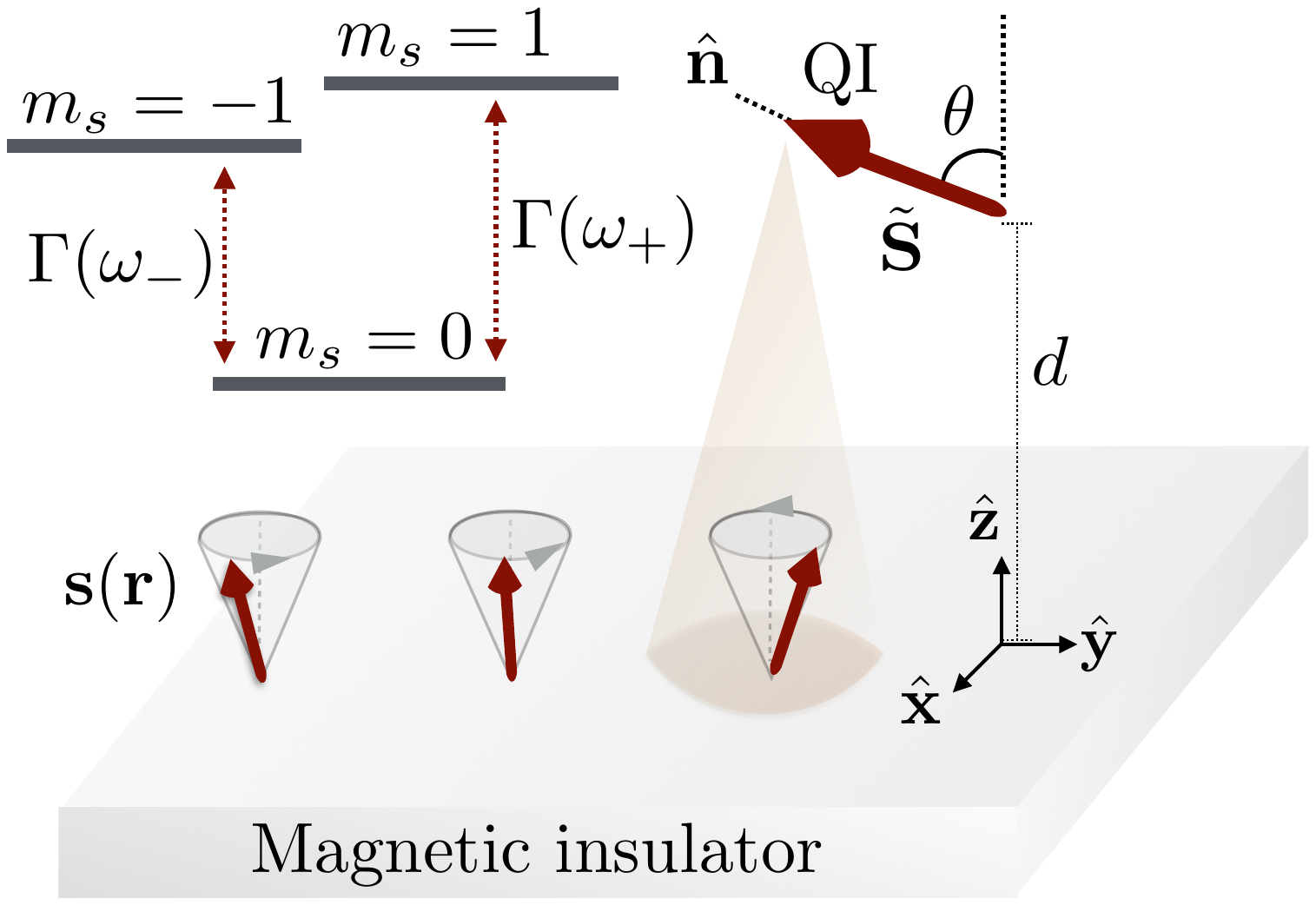}
\caption{ Setup for quantum-impurity relaxometry of a magnetic insulating system. The QI spin $\tilde{\mathbf{S}}$ is located at a height $d$ above the magnetic film and oriented along its anisotropy axis $\hat{\mathbf{n}}$, with $ \hat{\mathbf{z}} \cdot \hat{\mathbf{n}}=\cos\theta$. The coordinate system has the $xy$ plane placed on the magnetic film, with origin  aligned with the QI position.  The interactions between the QI spin and the local spin density $\textbf{s}(\mathbf{r})$ of the magnetic film induce  QI transitions between the spin states $m_{s}=0 \leftrightarrow \pm 1$  with energy loss or gain of $\hbar \omega_{\pm}$ at the rate $\Gamma(\omega_{\pm})$. }
\label{Fig2}
\end{figure}

\textit{Model.} In this work, we focus, for simplicity, on axially-symmetric magnetic insulating films with approximate \textit{U}(1) symmetry and a strong collinear order. In such systems,  the net spin parallel to the magnetic symmetry axis is (approximately) conserved. 
The spin-wave dynamics can then be described in terms of transport of the conserved component of the spin density, and, morever, we can introduce a well-defined magnon chemical potential \cite{Halperin1969,Bender2012,flebus2018}. In the setup we envision, illustrated in Fig.~\ref{Fig2}, a quantum-impurity spin $\tilde{\mathbf{S}}$, with $|\tilde{\mathbf{S}}|=1$, is placed at a height $d$ above a magnetic film and possesses its own anisotropy axis $\hat{\mathbf{n}}$, with $ \hat{\mathbf{z}} \cdot \hat{\mathbf{n}}=\cos\theta$.  The local spin density  $\mathbf{s}(\textbf{r})$ of the magnetic film generates a stray field  $\mathbf{B}(\mathbf{r}_{0})=  \gamma \int d^{2} \mathbf{r} \; \mathcal{D}(\mathbf{r},\mathbf{r}_{0}) \mathbf{s}(\mathbf{r})$ at the QI position $\mathbf{r}_{0}=(0,0,d)$, where
  $\gamma$ is  the gyromagnetic ratio of the film and $\mathcal{D}$ the tensorial magnetostatic Green's function \cite{Guslienko2011}.
Up to leading order in perturbation theory, the Zeeman coupling between the quantum-impurity spin and the stray field induces QI transitions between the spin states $m_{s}=0 \leftrightarrow \pm 1$ at the resonance frequency $\omega_{\pm}$.   We find the corresponding transition rate as
\begin{align}
\Gamma (\omega_{\pm})=f(\theta) \int^{\infty}_{0} dk \;  k^{3} e^{-2kd}  \left[   C_{\bot}(k, \omega_{\pm}) + 2 C_{\parallel}(k, \omega_{\pm}) \right]\,,
 \label{relaxationrate}
\end{align}
with $f(\theta)=  (\gamma \tilde{\gamma})^2  ( 5 -\cos2\theta)/32 \pi$, where $\tilde{\gamma}$ is the QI gyromagnetic ratio.  Here, $C_{\bot (\parallel)}(k, \omega_{\pm})$ is the real part of the Fourier transform of the spin-spin correlator  $C_{\bot (\parallel)}
(\mathbf{r}_{i},\mathbf{r}_{j}; t)=\{ s^{+(z)}(\mathbf{r}_{i},t), s^{-(z)}(\mathbf{r}_{j},0) \}$, which describes magnetic noise transverse 
(longitudinal)  to the magnetic symmetry axis $\hat{\mathbf{z}}$, i.e., to the equilibrium orientation of the order parameter. 
Invoking the Holstein-Primakoff transformation~\cite{Holstein1940}, i.e., $s^{+} \propto a^{\dagger} $ and $s_{z} \propto a^{\dagger} a$, with $a^{\dagger}$ ($a$) being the magnon creation (annihilation) operator,  one can see that the transverse and longitudinal noises emerge from, respectively, one- and two-magnon processes. In the following,  we assume the quantum-impurity frequency to lie within the magnetic gap, such that, in the limit of vanishing Gilbert damping~\cite{footnotedamping}, only two-magnon processes can take place, i.e., $C_{\bot}(\omega_{\pm})=0$.

\begin{figure}[t!]
\includegraphics[width=0.9\linewidth]{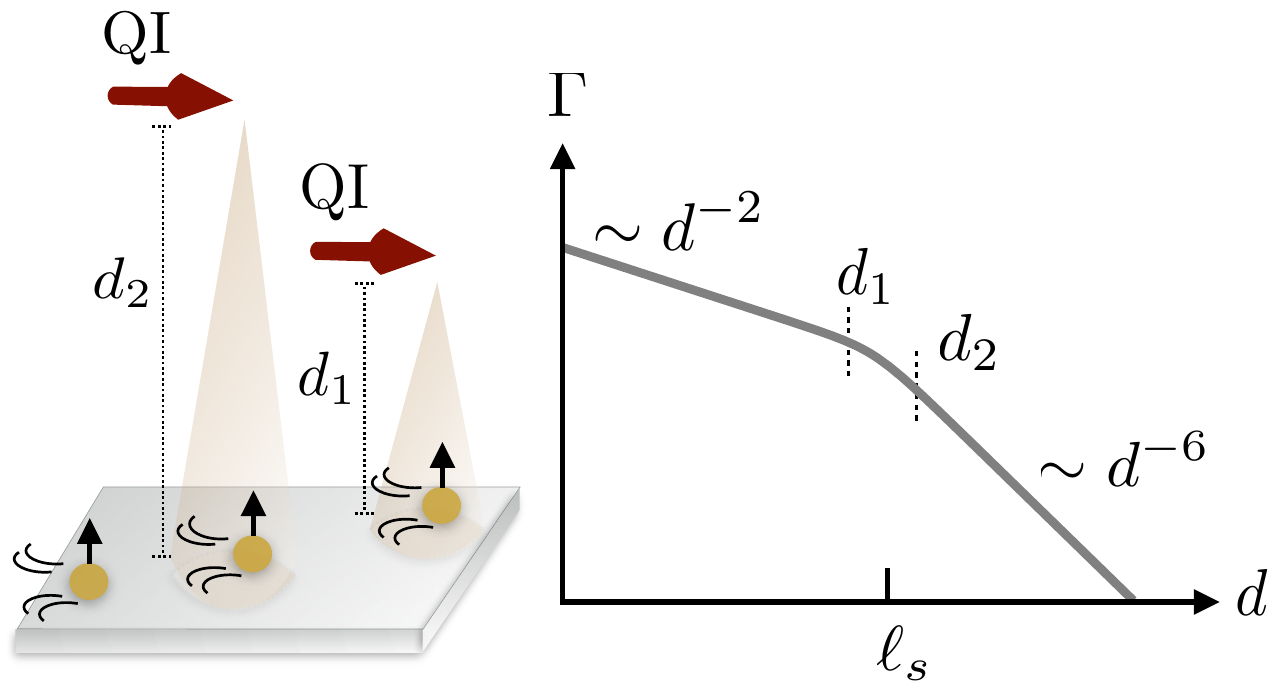}
\caption{Measurement of the spin diffusion length $\ell_{s}$. By varying the distance $d$ between the magnetic film and the quantum impurity, at low frequency one can find the cross-over region between two limiting behaviors of the QI relaxation rate $\Gamma$, i.e.,   $\Gamma \sim d^{-2}$ and $\Gamma \sim d^{-6}$. Within this region, measuring the relaxation rate at two different heights, $d_{1}$ and $d_{2}$, leads to an estimate for the spin diffusion length $\ell_{s}$.}
\label{Fig3}
\end{figure}

\textit{Diffusive transport properties via two-magnon noise.} The longitudinal noise, $C_{\parallel}$,  can be related to the imaginary part, $\chi^{''}_{\parallel}$, of the longitudinal spin susceptibility  via the fluctuation-dissipation 
theorem~\cite{kubo1966}, i.e., $C_{\parallel}(\omega,k)=\coth (\beta \hbar  \omega/ 2) \chi^{''}_{\parallel}(k,\omega)$, with $\beta=1/k_{B} T$ and $k_{B}$ being the Boltzmann constant. Thus the two-magnon driven QI relaxation rate is fully determined by the longitudinal spin susceptibility of the magnetic system. The latter depends on the pertinent spin transport regime, and it can be obtained by inverting the spin transport equations. As an experimentally-relevant example, here we consider a weakly-interacting magnon system, whose spin density dynamics can be treated as diffusive at wavelengths larger than the magnon mean free path $\ell_{\text{mp}}$, i.e., 
\begin{align}
\partial_{t} s_{z} + \boldsymbol{\nabla} \cdot \mathbf{j}_{s}=-\frac{1}{\tau_{s}}  s_{z} \,.
\end{align}
Here, we have introduced the spin-relaxation time $\tau_{s}$ and the spin current $\mathbf{j}_{s}=-\sigma \boldsymbol{\nabla} \mu\,$, where  $\sigma$ is the magnon spin conductivity,  $\mu=   \chi^{-1} s_{z} -\gamma H
$  the chemical potential,  $\chi$ the static uniform longitudinal susceptibility  and
 $H$  an external magnetic field. Introducing the diffusion coefficient $D=\sigma / \chi$, the imaginary part of the dynamical longitudinal spin susceptibility can be written as
\begin{align}
\chi^{''}_{\parallel}(k,\omega)= \frac{ \chi \omega D k^2 }{ (D k^2+ 1/\tau_{s})^2 + \omega^2}\,.
\label{114}
\end{align}
One might notice that, in Eq.~(\ref{relaxationrate}), the filtering function $k^{3} e^{-2kd}$, introduced by dipolar interactions, is peaked around the wave vector $k \sim 1/d$ : contributions to Eq.~(\ref{relaxationrate}) from smaller wave vectors are algebraically suppressed as they have limited phase space, while the ones at larger wavevectors are exponentially suppressed due to the self-averaging of short-wavelength fluctuations \cite{Agarwal2017}. This allows us to approximate $\chi_{\parallel}^{''} (k) \sim \chi_{\parallel}^{''} (1/d)$.
For  $\beta \omega \ll 1$, the quantum-impurity relaxation rate reads as 
\begin{align}
 \Gamma(\omega) \sim f(\theta) \frac{\hbar \chi}{ \beta D d^2}  \frac{1}{\left[ 1 +\left( \frac{d}{\ell_{s}} \right)^2 \right]^2 + \big( \frac{\omega d^2}{D} \big)^2}\,.
\label{86}
\end{align}
Measuring the QI relaxation rate while varying the distance between the quantum impurity and the magnetic film should then unveil the region over which a diffusive description of transport holds, according to Eq.~(\ref{86}), as well as the wavelength at which it starts breaking down.  Equation~(\ref{86}) shows that the relaxation rate increases with decreasing frequency, up to become constant, i.e., $ \Gamma \sim ( d+ d^{3}/\ell_{s})^{-2}$, for $\omega \ll D/d^2$. In this regime,  one can detect the region where $d \sim \ell_{s}$ as the cross-over region between $\Gamma \sim d^{-2}$ and $\Gamma \sim d^{-6}$, as depicted in Fig.~\ref{Fig3}. Within such region, we find that measuring the QI relaxation rates, $\Gamma(d_{1})$ and $\Gamma(d_{2})$, at two different distances, $d_{1}$ and $d_{2}$,  leads to an estimate for the spin diffusion length as 
\begin{align}
\ell^{2}_{s} \sim \frac{d^3_{1} \sqrt{\Gamma(d_{1})/\Gamma(d_{2})} - d^3_{2}}{d_{2} - d_{1} \sqrt{ \Gamma(d_{1})/\Gamma(d_{2}) }}\,.
\label{estimatediff}
\end{align} 
Since the spin-relaxation time $\tau_{s}$ and the susceptibility $\chi$ can be directly measured, one can use Eq.~(\ref{estimatediff}) to extract the magnon spin conductivity $\sigma$. Such  measurement, which could be performed by, e.g., embedding a NV center on a cantilever~\cite{Zhou,
Rondin2012}, would provide a direct probe of bulk spin transport properties, not marred by interfacial effects that affect conventional spin-transport experiments \cite{CornelissenPRB2016}.

\begin{figure}[t!]
\includegraphics[width=1\linewidth]{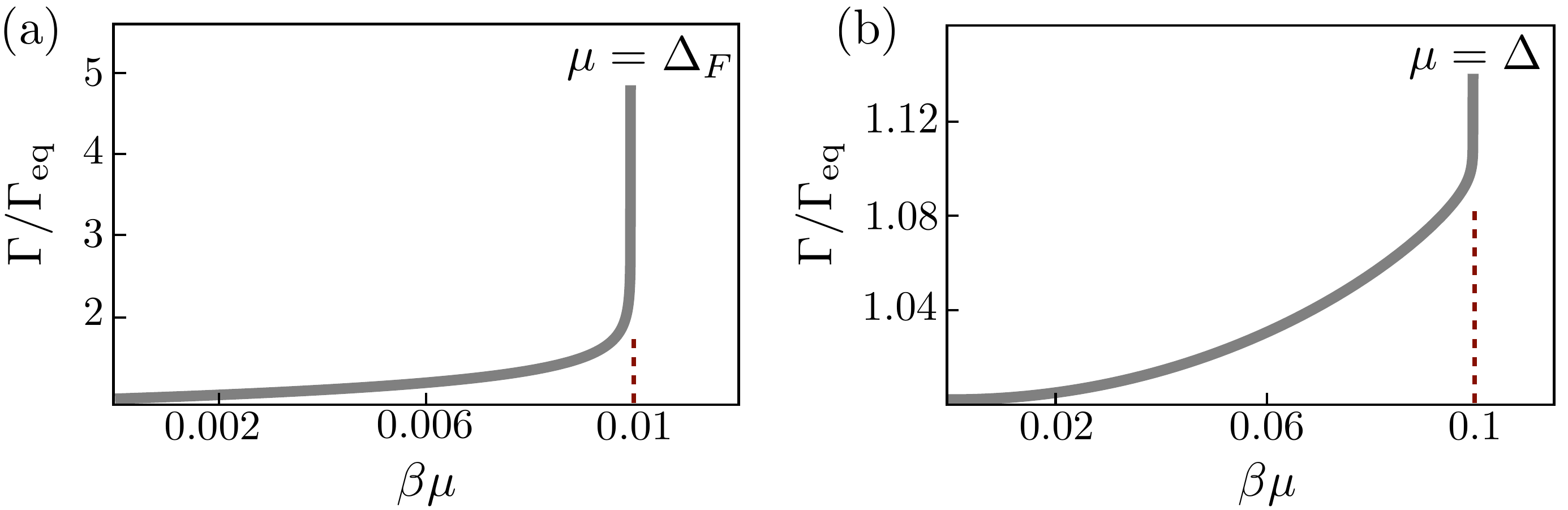}
\caption{QI relaxation rate as function of the magnon chemical potential for a quantum impurity interacting with (a) a ferromagnetic film with gap $\Delta_{F}=1$ K, at $T=100$ K; (b)  an antiferromagnetic film having gap $\Delta=1$ K, at $T=10$ K. The relaxation rate $\Gamma$ is normalized by its value $\Gamma_{\text{eq}}$ at equilibrium, i.e., for $\mu=0$.  }
\label{FigNoise}
\end{figure}

\textit{Magnon BEC via two-magnon noise.} 
As an example of detection, via two-magnon noise, of a dynamical phase transition, we focus on magnon Bose-Einstein condensation and, therefore, investigate the dependence of the two-magnon noise on the magnon chemical potential.  Our starting point is a general $\textit{U}(1)$-symmetric Hamiltonian
\begin{align}
\mathcal{H}_{m}=- J  \sum_{ \langle i \neq j \rangle}  \mathbf{S}_{i} \cdot \mathbf{S}_{j} + \gamma \sum_{i} \mathbf{S}_{i} \cdot \mathbf{H}  + \frac{K}{2} \sum_{i} (S^{z}_{i})^{2}\,,
\label{U1Hamiltonian} 
\end{align}
where  $\mathbf{S}_i$ is the dimensionless onsite spin at the site $\mathbf{r}_{i}$, $\mathbf{H}=H \mathbf{\hat{z}}$ is an uniform magnetic field, with $H>0$, $J$  the  exchange stiffness, and $K$ the constant governing the strength of the local  anisotropy. First, we consider a ferromagnetic system with easy-plane anisotropy, i.e., $J, K>0$. Introducing the Holstein-Primakoff transformation at leading order \cite{Holstein1940}, we truncate the resulting Hamiltonian up to quadratic order and Fourier transform it. Equation~(\ref{U1Hamiltonian}) is diagonalized by a magnon mode with chemical potential $\mu$ and dispersion  $\hbar \omega_{\mathbf{k}}=A k^2 + \Delta_{F}$, where $A \sim J S a^2_{0}$ is the spin stiffness,  $a_{0}$  the atomic spacing, and  $\Delta_{F}$  the ferromagnetic gap. 
In the continuum limit, the quantum-impurity spin couples to the coarsed-grained spin density (in physical units). 
For $d \ll \lambda_{T}$, with $\lambda_{T}$ being the magnon thermal de Broglie wavelength,  the dipolar kernel $\mathcal{D}(\mathbf{r}-\mathbf{r}_{0})$ can be approximated by a local coupling between the quantum impurity spin and the gradient of the longitudinal spin density, $s_{z}$, underneath it. Namely, for the magnetic field $\mathbf{B}(\mathbf{r}_{0})=\int d^{2}\mathbf{r} \sum_{i}\mathcal{D}_{3i}(\mathbf{r}-\mathbf{r}_{0}) s_{z}(\mathbf{r})$, we expand the longitudinal spin density in a Taylor series as $s_{z}(\mathbf{r}) \simeq s_{z}(0) + \sum_{\rho=x,y}\rho\partial_{\rho} s_{z}(0)$.  For $\hbar \omega \ll J (a_{0}/d)^2$ and $  \Delta_{F}, \mu \ll \beta^{-1}$, we obtain, setting $\theta=0$~\cite{thetachoice},
\begin{align}
\Gamma \approx \frac{   \hbar^3 ( \gamma \tilde{\gamma})^2}{\beta^2 A^3} \text{Log} \left[ \frac{1}{\beta ( \Delta_{F} -\mu)}\right]. 
\label{125}
\end{align}
Hence, while increasing the magnon chemical potential, which could be achieved via, e.g., microwave pumping~\cite{Du2017}, the two-magnon noise increases logaritmically and reaches its saturation value in correspondence of the precipitation of Bose-Einstein condensation, i.e., $\mu= \Delta_{F}$, as shown in Fig.~\ref{FigNoise}(a).
For an estimate of Eq.~(\ref{125}), we consider yttrium iron garnet (YIG), which is widely used in spintronic devices due to its long-range spin transport properties. Taking $|\gamma|=|\tilde{\gamma}|=2\mu_{B}/\hbar$, where $\mu_{B}$ is the Bohr magneton (in cgs units), and using typical YIG parameters~\cite{Bhagat}, for a film of thickness $t \sim 10$ nm and at temperature $T \sim 100$ K, we obtain a relaxation time $\Gamma^{-1} \sim 100$ ms. The latter is shorter than the intrinsic relaxation time of, e.g., NV centers~\cite{BarGill2013}, suggesting that the signal could be experimentally detected.

Next, we consider an antiferromagnetic system with easy-axis anisotropy, whose energetics can described by Eq.~(\ref{U1Hamiltonian}) setting   $K,J \rightarrow - K,-J $, while keeping $K,J>0$.
Introducing the Holstein-Primakoff transformation up to leading order and, consequently, a Bogoliubov transformation (see, e.g., Ref.~\cite{Rezende2016}), we can diagonalize the resulting Fourier transform of Eq.~(\ref{U1Hamiltonian}) in terms of two magnon eigenmodes, each one carrying spin angular momentum $\pm \hbar$, whose distribution function are characterized by chemical potential $\pm \mu$  and dispersion $\hbar \omega_{\mathbf{k}} \mp  \gamma H$ \cite{flebus2018}.  Here, we have introduced  $\hbar \omega_{\mathbf{k}}=\sqrt{\Delta^{2}+(c k)^2 }$, with $c \sim J S a_{0}$ being the spin-wave velocity and $\Delta$ the antiferromagnetic gap.  The quantum-impurity spin couples to the spin density (in physical units)  of both sublattices. For $\hbar \omega \ll J a_{0}/d$ and $ \Delta, \mu \ll \beta^{-1}$,  the transition rate reads as $\Gamma \sim 2 \hbar^3 [ J_{2}(\mu) + J_{2}(-\mu)] (\gamma \tilde{\gamma})^2 / \beta^5 c^6$, where $J_{a}(\mu)=\int^{1}_{\beta \Delta} x^a [x^2- (\beta \Delta)^2] (x-\beta \mu)^{-2}$ is a dimensionless integral and we have set $\mu \rightarrow \mu -\gamma H$. When the system approaches the condensation point, i.e.,  $\mu \rightarrow \Delta$, the integrand  displays a singularity at the bottom of the magnon dispersion, where the magnon dispersion is quadratic. Hence, one might expect the two-magnon relaxation rate to display a divergence analogous to Eq.~(\ref{125}). 
 Indeed,  we find that, for $\mu \rightarrow \Delta$, the relaxation rate can be written as
\begin{align}
\Gamma \approx \frac{2\hbar^3( \gamma \tilde{\gamma})^2}{  \beta^5 c^6} \left[ \frac{1}{3} + (\beta  \Delta)^3 \; \text{Log} \left[ \frac{1}{\beta ( \Delta -\mu)}\right] \right]\,.
\label{compensatedtwomagnon}
\end{align}
For an estimate of Eq.~(\ref{compensatedtwomagnon}), we consider the Heisenberg-like antiferromagnet $\text{RbMnF}_{3}$, due to its simple cubic structure and its small exchange interaction and magnetic gap~\cite{Coldea1998}. Taking $J \approx 70$ GHz, $a_{0} \approx 4 \; \dot{\text{A}}$, $S=5/2$,  $\Delta \approx 1$ K \cite{Coldea1998,Jiang2006}, and setting $T=10$ K, we find that   $\Gamma^{-1} \sim 10 \; \mu\text{s}$.  Figure~\ref{FigNoise}(b) shows that, while the contribution from the magnon branch with negative chemical potential lowers the growth of the overall relaxation rate, the onset of Bose-Einstein condensation is still signalled by a logarithmic singularity.
%  Finally, we consider an antiferromagnetic system that displays a slightly uncompensated spin density $s$, with $s/s_{0} \ll 1$, where $s$ is the saturated spin density.  For $\omega \ll J a_{0}/d$ and $ \Delta, \mu \ll \beta^{-1}$, we find the transition rate  as $\Gamma \sim  (\gamma \tilde{\gamma} \hbar)^2 /( \beta^2 c^3) (s/s_{0})^2 [ J_{-2}(\mu) + J_{-2}(-\mu)]$.
%%Making use of the previous parameters, we obtain a relaxation rate of $\Gamma \sim  \; \mu\text{s}^{-1}$ at equilibrium. In this case, the onset of magnon condensation leads to a 5-fold increase in the relaxation rate, as shown in Fig.~\ref{Fig3}(d) .

\textit{Discussion.} In this work, we find that two-magnon noise, which has been, heretofore, neglected, can be used as a direct probe of spin-wave bulk transport properties in magnetic insulators. With the growing interest in insulating systems with long-range spin transport capabilities, we propose quantum-impurity relaxometry as a direct probe of key quantities such as the spin-diffusion length, without a need to fabricate metal$|$insulator heterostructures. While we have explicitly focused on  diffusive spin waves, our framework can be straightforwardly extended to other transport regimes.
 Moreover, our results suggest that magnon Bose-Einstein condensation can be detected via two-magnon noise in both ferromagnetic and antiferromagnetic systems. Our findings can be readily tested experimentally in ferromagnetic insulators, such as YIG, and, most importantly, they open up new prospects for detecting magnon condensation, induced by, e.g., thermal gradients, in antiferromagnetic insulators~\cite{TserkovnyakPRB2016}. With its combined capabilities,  quantum-impurity relaxometry driven by two-magnon noise might also shed light on the spin-transport properties of systems in which thermal spin waves cohexist with a superfluid condensate of magnons~\cite{FlebusPRL2016}.

In this work, we focused on magnetic insulators, where, due to the lack of charge noise, we can directly relate the QI relaxation rates to one- and two-magnon processes. Our theory, however, can be applied also to conducting materials, in the regimes when the magnetostatic noise associated with spin-density fluctuations dominates over the electronic (Johnson-Nyquist) noise.

\begin{acknowledgments}
The authors thank T. Van Der Sar for insightful discussions and the International Institute of Physics in Natal, Brazil, where this work was initiated, for their generous hospitality. B.F. was supported by the Dutch Science Foundation (NWO) through a Rubicon grant and Y.T. by NSF under Grant No. DMR-1742928.
\end{acknowledgments}

%Geared towards collective processes in magnon transport.
%Future work should address the detection, via quantum-impurity relaxometry, of spin superfluid modes, which have been theoretically predicted in a variety of \textit{U}(1)-symmetric magnetic systems \cite{FlebusPRL2016,Takei2014,Kim2017} and up to now not yet experimentally observed.  Understanding the interactions between such modes and a quantum-impurity, alongside with the  characterization of the noise produced by the host magnetic system, could open new prospects to couple distant quantum impurities.


\begin{thebibliography}{99}

\bibitem{Taylor2008} J. M. Taylor, P. Cappellaro, L. Childress, L. Jiang, D. Budker, P. R. Hemmer, A. Yacoby, R. Walsworth, and M. D. Lukin,  Nat. Phys. \textbf{4}, 810–6 (2008).

\bibitem{Pham2011} L. M. Pham, D. Le Sage, P. L. Stanwix, T. K. Yeung, D. Glenn, A. Trifonov, P. Cappellaro, P. R. Hemmer, M. D. Lukin, H. Park, A. Yacoby, R. L. Walsworth, New J. Phys. \textbf{13}, 045021 (2011).

\bibitem{Hong2013} S. Hong, M. S. Grinolds, L. M. Pham, D. Le Sage, L. Luan, R. L. Walsworth, and A. Yacoby, MRS Bull. \textbf{38}, 155–61 (2013).

\bibitem{Grinolds2013} M. S. Grinolds, S. Hong, P. Maletinsky, L. Luan, M. D. Lukin, R. L. Walsworth, A. Yacoby, Nature Phys. \textbf{9}, 215-219 (2013).
%Nanoscale magnetic imaging of a single electron spin under ambient conditions

\bibitem{kubo1966} R. Kubo, Rep. Prog. Phys. \textbf{29}, 255 (1966).

\bibitem{Agarwal2017} K. Agarwal, R. Schmidt, B. Halperin, V. Oganesyan, G. Zarand, M. D. Lukin, and E. Demler, Phys. Rev. B \textbf{95}, 155107 (2017).

\bibitem{Du2017} C. Du, T. Van der Sar, T. X. Zhou, P. Upadhyaya, F. Casola, H. Zhang, M. C. Onbasli, C. A. Ross, R. L. Walsworth, Y. Tserkovnyak, and A. Yacoby, Science \textbf{357} (6347), 195-198  (2017).

\bibitem{CornelissenNature2015} L. J. Cornelissen, J. Liu, R. A. Duine, J. Ben Youssef, and B. J. van Wees, Nat. Phys. \textbf{11}, 1022 (2015).


\bibitem{CornelissenPRB2016} L. J. Cornelissen, K. J. H. Peters, G. E. W. Bauer, R. A. Duine, and B. J. van Wees, Phys. Rev. B \textbf{94}, 014412 (2016).

\bibitem{ChumakAPL2012} A. V. Chumak, V. I. Vasyuchka, A. A. Serga, and B. Hillebrands, Nature Physics \textbf{11}, 453 (2015).


\bibitem{DerSar2015} T. van der Sar, F. Casola, R. Walsworth, and A. Yacoby,  Nat. Commun. \textbf{6}, 7886 (2015).

\bibitem{Baltz2018} V. Baltz, A. Manchon, M. Tsoi, T. Moriyama, T. Ono, and Y. Tserkovnyak
Rev. Mod. Phys. \textbf{90}, 015005. 

\bibitem{Takei2014} S. Takei, B. I. Halperin, A. Yacoby, and Y. Tserkovnyak, Phys. Rev. B \textbf{90}, 094408 (2014).

\bibitem{Halperin1969} B. I. Halperin and P. C. Hohenberg, Phys. Rev. \textbf{188}, 898 (1969).

\bibitem{Bender2012} S. A. Bender, R. A. Duine, and Y. Tserkovnyak, Phys. Rev. Lett. \textbf{108}, 246601 (2012). 

\bibitem{flebus2018} B. Flebus \textit{et al.}, in preparation.
 \bibitem{Guslienko2011} K. Y. Guslienko, and A. N. Slavin, J. Magn. Magn. Mater. \textbf{323}, 2418–2424 (2011).

\bibitem{Holstein1940} T. Holstein and H. Primakoff, Phys. Rev. \textbf{58}, 1098 (1940).

\bibitem{footnotedamping} Relaxation of spin dynamics gives rise to a broadening $\sim 1/\tau_{s}$ in the FMR (AFMR) absorption spectrum, with $\tau_{s}$ being the spin-relaxation time. The single-magnon noise then decays rapidly within the subgap region, essentially vanishing at frequencies $\omega \lesssim \Delta - 1/\tau_{s}$.

\bibitem{Zhou} T. X. Zhou, R. J. Stohr, and  A. Yacoby, Appl. Phys. Lett. \textbf{111}, 163106 (2017).

\bibitem{Rondin2012}
L. Rondin, J.-P. Tetienne, P. Spinicelli, C. Dal Savio, K. Karrai, G. Dantelle, A. Thiaville, S. Rohart, J.-F. Roch, V. Jacques, Appl. Phys. Lett. \textbf{100}, 153118 (2012). 


\bibitem{thetachoice} Note that a different choice for the anisotropy axis orientation $\theta$ would only affect the overall multiplication factor.

\bibitem{Bhagat} S. Bhagat, H. Lessoff, C. Vittoria, and C. Guenzer, Phys. Status Solidi \textbf{20}, 731 (1973).

\bibitem{BarGill2013} S. Amashm, K. MacLean, I. P. Radu, D. M. Zumbuhl, M. A. Kastner, M. O. Hanson, and A. C. Gossard, Phys. Rev. Lett. \textbf{100}, 046803 (2008).

\bibitem{Rezende2016} S. M. Rezende, R. L. Rodriguez-Suarez, and A. Azevedo, 
Phys. Rev. B \textbf{93}, 054412 (2016).

\bibitem{Coldea1998} R. Coldea, R. A. Cowley, T. G. Perring, D. F. McMorrow, and
B. Roessli, Phys. Rev. B \textbf{57}, 5281 (1998).

\bibitem{Jiang2006} L. Jiang, J. Guo, H. Liu, M. Zhu, X. Zhou, P. Wu, and C. Li, J. Phys. Chem. Solids \textbf{67}, 1531 (2006).

\bibitem{TserkovnyakPRB2016} Y. Tserkovnyak, S. A. Bender, R. A. Duine, and B. Flebus, Phys. Rev. B \textbf{93}, 100402 (2016).

\bibitem{FlebusPRL2016} B. Flebus, S. A. Bender, Y. Tserkovnyak, and R. A. Duine, Phys. Rev. Lett. \textbf{116}, 117201 (2016).

\end{thebibliography}
\end{document}